\documentclass[10pt,twocolumn]{article}

\usepackage{graphicx} 
\graphicspath{{figures}} 
\usepackage[font=small,labelfont=bf]{caption} 

\usepackage{xcolor}
\definecolor{black}{HTML}{212427}
\definecolor{blue}{HTML}{0563C1}
\definecolor{brightred}{HTML}{FF0000} 

\usepackage{hyperref,nameref}
\hypersetup{colorlinks=true,allcolors=blue}
\newcommand{\rref}[2]{\hyperref[#1]{\ref{#1}#2}} 

\color{black}
\usepackage[margin=0.67in]{geometry} 
\usepackage[switch]{lineno} 

\usepackage{fancyhdr,lastpage} 
\usepackage{titlesec}
\titleformat{\section}{\bfseries}{}{1mm}{}
\titleformat{\subsection}{\bfseries}{\thesubsection}{1mm}{}
\titlespacing{\section}{0pt}{10pt}{0pt}
\titlespacing*{\section}{-3pt}{\baselineskip}{0pt}
\titlespacing{\subsection}{0pt}{10pt}{0pt}

\usepackage{amsmath,amssymb}
\renewcommand{\v}[1]{\boldsymbol{\mathbf{#1}}} 
\newcommand{\Ang}[0]{\mathring{\mathrm{A}}} 
\renewcommand{\l}[0]{\left} 
\renewcommand{\r}[0]{\right} 
\let\f=\frac 
\renewcommand{\t}[1]{\text{#1}} 

\newcommand{\rev}[1]{{#1}} 

\usepackage[
  backend=biber,        
  autocite=superscript, 
  sorting=none,         
  style=numeric-comp,   
  maxnames=100,         
  minnames=100,         
  giveninits=false,     
  autopunct=false,      
  doi=true,             
  hyperref=true         
]{biblatex}
\DeclareFieldFormat{labelnumberwidth}{\;[#1]\!\!\!\!} 
\renewbibmacro{in:}{} 
\AtNextBibliography{\small} 
\AtEveryBibitem{\clearfield{pages}} 
\AtEveryBibitem{\clearfield{volume}} 
\AtEveryBibitem{\clearfield{number}} 

\begin{document}

\twocolumn[
  \begin{center}
	\large
	\textbf{Quantifying chemical short-range order in metallic alloys}
  \end{center}

  Killian Sheriff$^1${\footnotemark[1]},
  Yifan Cao$^1${\footnotemark[1]},
  Tess Smidt$^2$, and
  Rodrigo Freitas$^1${\footnotemark[2]} \\
  $^1$\textit{\small Department of Materials Science and Engineering, Massachusetts Institute of Technology, Cambridge, MA, USA} \\
  $^2$\textit{\small Department of Electrical Engineering and Computer Science, Massachusetts Institute of Technology, Cambridge, MA, USA} \\

  {\small Dated: \today}

  \vspace{-0.15cm}
  \begin{center}
	\textbf{Abstract}
  \end{center}
  \vspace{-0.35cm}

  Metallic alloys often form phases --- known as solid solutions --- in which chemical elements are spread out on the same crystal lattice in an almost random manner. The tendency of certain chemical motifs to be more common than others is known as chemical short-range order (SRO) and it has received substantial consideration in alloys with multiple chemical elements present in large concentrations due to their extreme configurational complexity (e.g., high-entropy alloys). Short-range order renders solid solutions ``slightly less random than completely random'', which is a physically intuitive picture, but not easily quantifiable due to the sheer number of possible chemical motifs and their subtle spatial distribution on the lattice. Here we present a multiscale method to predict and quantify the SRO state of an alloy with atomic resolution, incorporating machine learning techniques to bridge the gap between electronic-structure calculations and the characteristic length scale of SRO. The result is an approach capable of predicting SRO length scale in agreement with experimental measurements while comprehensively correlating SRO with fundamental quantities such as local lattice distortions. This work advances the quantitative understanding of solid-solution phases, paving the way for the rigorous incorporation \rev{of SRO length scales} into predictive mechanical and thermodynamic models.

  \vspace{0.4cm}
]
{
  \footnotetext[1]{These authors contributed equally to this work.}
  \footnotetext[2]{Corresponding author (\texttt{rodrigof@mit.edu}).}
}

\noindent Short-range order exists because low-energy chemical motifs are favored in thermally-equilibrated solid solutions, driving them away from complete randomness. This phenomena occurs widely\autocite{mark_mrs} in ceramics materials\autocite{HEC_1,HEC_2,HEC_3} and --- the focus of this article --- metals\autocite{HEA,mark_mrs}. Quantifying this tendency is a tantalizing goal because SRO effectively functions as the background against which microstructural evolution occurs. Naturally, it has been broadly suggested that various chemistry--microstructure relationships are affected by SRO, and that its manipulation could be employed with the purpose of designing materials properties and performance\autocite{andy,SRO_dislocations,SRO_SFE,SRO_radiation,penghui,SRO_simulation}. Yet, even in completely random solid solutions one is bound to encounter random chemical fluctuations that resemble SRO. These random fluctuations obfuscate SRO in thermally equilibrated solid solutions, making SRO identification challenging\autocite{jim_null,EXAFS_SRO,tem_challenge,resistivity_easo} and entangling their effects on materials properties.

Indirect evidence has long been employed to establish SRO existence\autocite{SRO_indirect,resistivity_easo}, including incomplete quantitative metrics of SRO, i.e., quantities that serve as indicators of SRO but do not capture the entire complexity of chemical motifs and their spatial distribution\autocite{WC_complete,fontaine,khachaturyan,MB_WC}. Unequivocal quantification of SRO requires direct atomic-scale characterization, which is a difficult task to be performed experimentally --- as indicated by various recent reports\autocite{EXAFS_SRO,tem_challenge,resistivity_easo,APT_CoCrNi,diffuse_origin}. Fundamentally, the lack of atomic scattering contrast among transition metals hinder direct SRO identification using electron or x-ray diffraction. 

Quantifying SRO becomes an even more formidable problem in the space of high-entropy alloys because the presence of multiple chemical elements in similarly large concentrations enables tremendous flexibility in expressing SRO\autocite{EXAFS_SRO,WC_complete,fontaine}. While thought-provoking observations have been made about the role of SRO in various chemistry--microstructure interactions, a comprehensive framework for systematically connecting chemistry to microstructural evolution is still lacking. This absence speaks of a considerable challenge in quantifying SRO that is beyond the capability of current approaches. Here we remedy this by developing a predictive multiscale methodology for the quantification of SRO from atomic-scale data where the complexity of local chemical motifs is accounted for in its entirety --- even in the case of high-entropy alloys. The result is an approach that operates at the length scales characteristic of SRO and is capable of predicting SRO length scales in good agreement with experimental observations.

\section{Predictive calculations at appropriate length scales}

Quantification of SRO through computational efforts rely substantially on the physical fidelity of the underlying atomistic model. Here we employ \rev{spin-polarized} density-functional theory (DFT) calculations in order to capture \rev{magnetic contributions\autocite{flynn_PNAS} and other} nuances of the energetics of chemical bonding that ultimately lead to the existence of SRO. The target alloy system chosen is the solid-solution phase of CrCoNi; a paradigmatic face-centered cubic high-entropy alloy that has been widely investigated\autocite{SRO_radiation,SRO_SFE,andy,flynn_PNAS,exafs,tamm,SRO_dislocations,ma_NiCoCr}.

\begin{figure}[!tb]
  \centering
  \includegraphics{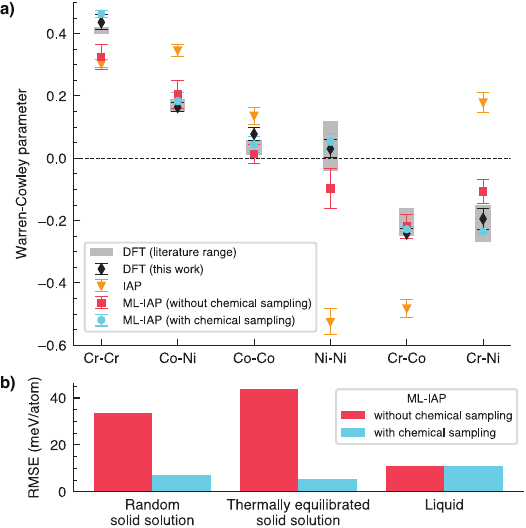}
  \caption{\label{figure_1} \textbf{Capturing chemical SRO with ML-IAPs.} Training a ML-IAP with extensive sampling of the chemical-ordering space (shown in blue) resulted in a SRO behavior with better agreement with DFT calculations than other popular approaches in the literature, including a traditional IAP\autocite{SRO_dislocations} (shown in orange) and a state-of-the-art approach for training ML-IAPs for solid solutions\autocite{SPO_MLIAP_1,SPO_MLIAP_2} that does not perform chemical sampling (shown in red). \textbf{a)} Warren-Cowley parameters at 500\,K. Error bars are the standard error from the mean obtained from five independent simulations. \textbf{b)} Energy root-mean-square error on independent test sets under thermodynamic conditions not included in the training of either ML-IAP.}
\end{figure}

Fundamentally, SRO is still most often equated to the Warren-Cowley (WC) parameters\autocite{WC_1,WC_2}:
\begin{equation}
  \label{eq:WC}
  \alpha_\t{AB} = 1 - \f{p(\t{A}|\t{B})}{c_\t{A}},
\end{equation}
where A and B indicate chemical elements, $p(\t{A}|\t{B})$ is the conditional probability of finding an element A at a nearest neighbor site of an element B, and $c_\t{A}$ is the average concentration of element A in the alloy. Figure \rref{figure_1}{a} shows CrCoNi's six WC parameters at 500\,K as evaluated with Monte Carlo DFT calculations, alongside with the range of values reported in the literature. While experimental evaluation of WC parameters for CrCoNi is not available, the calculated results are consistent with extended x-ray adsorption fine structure measurements\autocite{exafs} indicating, for example, that Cr-Cr bonds are unfavorable --- an observation that has since been confirmed by other techniques\autocite{APT_CoCrNi,ma_NiCoCr} and elucidated by physical arguments\autocite{flynn_PNAS,tamm,doug_jim,finnis}.

Experimental evidence\autocite{doug_jim,andy,ma,ma_NiCoCr,jim_cem,resistivity_easo} suggests that SRO length scale can be as large as 2\,nm, which makes proper statistical sampling of SRO though DFT-based calculations impossible. A size convergence analysis in Supplementary Section 1 confirms that system sizes typical of DFT calculations are indeed not converged with respect to SRO. Interatomic potentials (IAPs) are computationally inexpensive approaches often employed to circumvent the length-scale limitation of DFT. Yet, the only IAP for CrCoNi available in the literature\autocite{SRO_dislocations} is unable to reproduce the WC parameters --- not even qualitatively --- despite being frequently employed to investigate SRO effects (e.g., ref.~\cite{penghui}). For example, fig.~\rref{figure_1}{a} shows that this IAP predicts strong Ni-Ni attraction while DFT predicts a mild repulsion. To address this shortcoming we turn to machine learning (ML) IAPs. We adopted a popular strategy\autocite{SPO_MLIAP_1,SPO_MLIAP_2} for training ML-IAPs for solid solutions that employs a state-of-the-art ML model with better performance than other models in an independent assessment of various ML-IAPs\autocite{SPO_MLIAP_performance}. Figure \rref{figure_1}{a} shows that this approach --- labeled ``ML-IAP (without chemical sampling)'' --- is a marked improvement over the IAP, but still falls short of DFT predictions.

Excellent agreement with DFT predictions was obtained by developing an entirely new training approach centered around the extensive sampling of the chemical-ordering space, i.e., including chemical configurations ranging from random to thermally equilibrated. This approach --- shown as ``ML-IAP (with chemical sampling)'' in fig.~\ref{figure_1} --- also generalized much better to an independent test set under thermodynamic conditions not included in the training of either ML-IAPs (fig.~\rref{figure_1}{b}) without compromising on the accuracy of unrelated phases (represented in fig.~\rref{figure_1}{b} by the liquid phase). An analysis of the performance of this ML-IAP beyond WC parameters (i.e., using local chemical motifs) is provided in Supplementary Section 2.

Considering the substantial improvement over state-of-the-art established in fig.~\ref{figure_1}, we assert this to be the first approach to enable predictive calculations of SRO at appropriate length scales. Equipped with this new capability we set out next to quantify SRO from atomic-scale data.

\section{Short-range order representation and metric}

\begin{figure*}[!bt]
  \centering
  \includegraphics{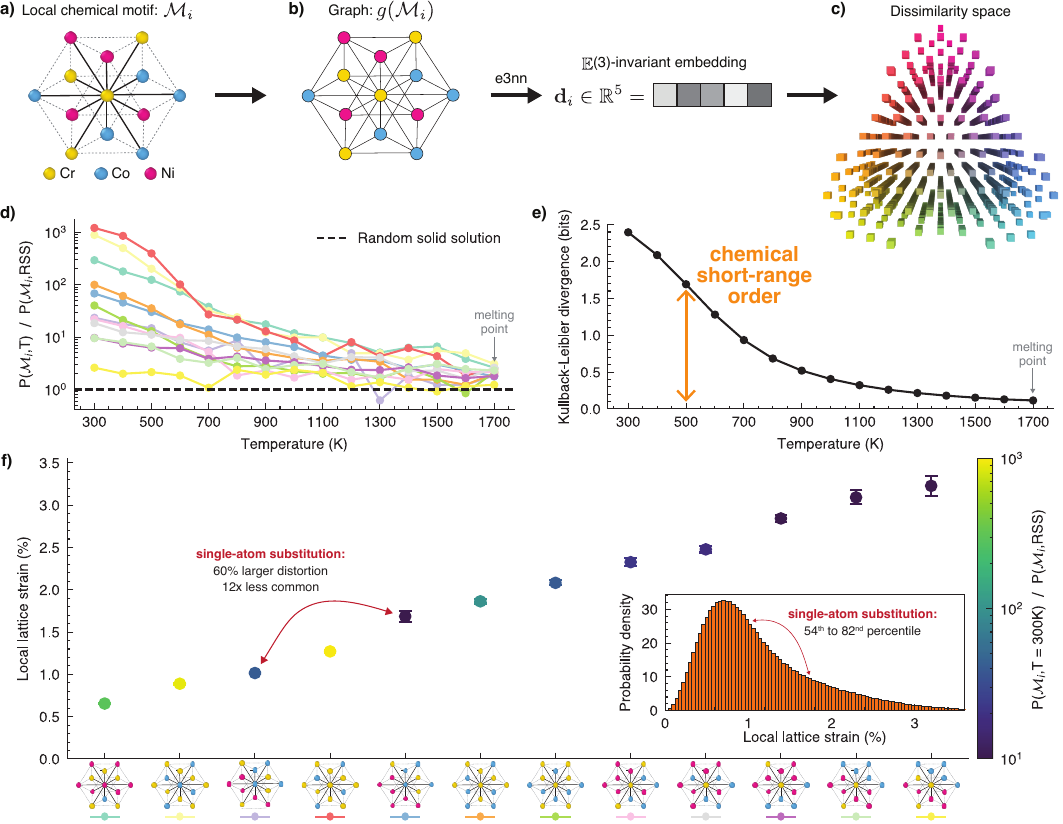}
  \caption{\textbf{Representing and quantifying SRO.} \textbf{a)} Local chemical motifs ($\mathcal{M}_i$) --- defined by an atom and its nearest neighbors --- characterize the chemical bond environment experienced by the central atom. \textbf{b)} A graph representation of local chemical motifs $g(\mathcal{M}_i)$ is employed to map physically equivalent chemical bond environments to the same embedding $\v{d}_i \in \mathbb{R}^5$ using a neural network that is equivariant to Euclidean $\mathbb{E}$(3) symmetry (e3nn). \textbf{c)} Space of dissimilarity among all 12\,111 unique motifs with a Cr central atom (the space for all 36\,333 unique motifs is five dimensional). \textbf{d)} The probability of observing representative motifs in thermal equilibrium, $P(\mathcal{M}_i, T)$, converges continuously towards the random solid-solution probability, $P(\mathcal{M}_i, \t{RSS})$. The motif corresponding to each color is shown at the bottom of fig.~\rref{figure_2}{f}. \textbf{e)} The Kullback-Leibler divergence ($D_\t{KL}$) of $P(\mathcal{M}, T)$ from $P(\mathcal{M}, \t{RSS})$ collects all of the information about local chemical motifs into a single metric that quantifies the amount of order in the system. \textbf{f)} Quantification of complex connection between local chemistry and local lattice distortion. The inset shows the density distribution of local lattice distortions at 300\,K. Motifs are shown in order of increasing distortion to facilitate visualization. Error bars are the standard error from the mean.}
  \label{figure_2}
\end{figure*}

The smallest building block for construction of a complete representation of SRO consists of an atom and its local chemical bond environment as defined by its nearest neighbors. A representative case of this construct, which we refer to as a \textit{local chemical motif}, is illustrated in fig.~\rref{figure_2}{a} for the face-centered cubic CrCoNi alloy. While there are a total of $3 \times 3^{12} = 1\,594\,323$ possible local chemical motifs such as this one, many of them lead to physically equivalent local chemical bond environments for the central atom. More rigorously, any two motifs are equivalent if they can be related to each other by rotations, inversions, or translations --- a set of operations that together are known as Euclidean symmetry or $\mathbb{E}$(3) group. Using a group theory approach known as Polya enumeration theorem\autocite{polya} we determined analytically that there are only 36\,333 unique chemical motifs --- represented here by $\mathcal{M}_i$, with $i = 1, 2, ...\,, 36\,333$. Supplementary Section 4 provides a detailed description of the application of Polya's enumeration theory to the counting of unique motifs.

The tendency of certain motifs to be more common than others can be quantified by the probability density $P(\mathcal{M},T)$, which carries fundamental statistical information defining the SRO state at temperature $T$. Because of this we would like to identify and count motifs in atomic-scale data, but Polya's enumeration theorem does not provide a practical way to classify an arbitrary motif according to its symmetry. In order to accomplish this we turn to a class of graph neural networks that are naturally equivariant to $\mathbb{E}$(3) symmetries, namely Euclidean neural networks (e3nn)\autocite{e3nn_1,e3nn_2,tfn,cgn,cnn}.

We start by representing local chemical motifs $\mathcal{M}_i$ in a graph form (fig.~\rref{figure_2}{b}): $g(\mathcal{M}_i)$. Classifying such graph according to its symmetry requires performing numerous graph comparisons in order to determine isomorphisms, which cannot be generally solved in polynomial time (i.e., this is a nondeterministic polynomial problem --- or NP problem). \rev{The need for this brute-force classification of each graph is circumvented by employing a randomly initialized e3nn to create an embedding representation $\v{d}_i \in \mathbb{R}^5$ of $g(\mathcal{M}_i)$, as illustrated in fig.~\rref{figure_2}{b} (see methods section for a complete description of network architecture). This representation is capable of discriminating between different graphs, i.e., $\v{d}_i = \v{d}_j$ if $g(\mathcal{M}_i)$ and $g(\mathcal{M}_j)$ are isomorphic, otherwise $\v{d}_i \neq \v{d}_j$. This capability originates from e3nn's ability to capture and encode intra-graph relationships and topology, effectively functioning as a symmetry compiler\autocite{e3nn_4,xu2019powerful} (which can be more rigorously explained by the similarities between e3nn's message passing algorithm and the Weisfeiler-Lehman graph-isomorphism test\autocite{WL_1,WL_2}). The random initialization is employed to maximize the influence of each neural-network weight\autocite{e3nn_4}.} Additionally, by employing e3nn we are guaranteeing $\v{d}_i$ to account for all $\mathbb{E}(3)$ symmetries and all subgroups, i.e., physically equivalent local chemical motifs are mapped to the same $\v{d}_i$. Application of this approach to the set of all 1\,594\,323 possible chemical motifs identifies only 36\,333 unique local chemical motifs, \rev{i.e., 36\,333 unique $\v{d}_i$ up to 8 significant digits,} confirming the analytical result of Polya's enumeration theorem with an approach that is computationally viable for application to large atomistic simulations\rev{, processing of $1.2\times 10^6$ atoms per hour in a single CPU core with a Apple Silicon M1 processor, or $63 \times 10^6$ atoms per hour on a NVIDIA V100s GPU. This approach has been generalized to different lattice structures (body-centered cubic and hexagonal close-packed) and to up to five chemical elements.}

Equipped with this approach we quantify how often motifs are observed in thermally equilibrated solid solutions, $P(\mathcal{M}, T)$, compared to a random solid solution $P(\mathcal{M}, \t{RSS})$. The result for representative motifs is shown in fig.~\rref{figure_2}{d}, where it can be observed that certain motifs are three orders of magnitude more common in equilibrium than in the random case --- an impressive expression of chemical SRO. Moreover, all motifs seem to continuously converge towards the random solid solution probability in the limit of high temperatures, which is a sensible physical behavior to expect based on statistical mechanics.

Figure \rref{figure_2}{d} illustrates how the approach introduced in figs.~\rref{figure_2}{a} and \rref{figure_2}{b} breaks the system down to its smallest SRO elements. However, one would prefer to have such information collected into a single metric that quantifies the amount of order in the system. Configurational entropy is a physically rigorous quantity that would be ideally suited for this task; \rev{while} the effect of SRO on configurational entropy can be accounted for with \rev{well-established cluster-based methods\autocite{S_config}}, such \rev{approaches} become inaccurate as SRO increases \rev{and require increasingly larger clusters}. Instead, we \rev{introduce an approach capable of describing any amount of SRO with two different components: (i) the tendency of certain chemical motifs $\mathcal{M}_i$ to be more common than others and (ii) how the motifs are organized in space (introduced below in the section ``Short-range order length scale''). To capture the tendency of certain motifs to be more common we turn to a generalized definition of entropy to probability distributions such as $P(\mathcal{M},T)$, namely the Shannon entropy}:
\begin{flalign}
  \label{eq:KL}
  D_\t{KL} \Big[ P(\mathcal{M},T) \; || \; P(\mathcal{M},\t{RSS}) \Big] = && \\
  \sum_{i=1}^{36\,333} P(\mathcal{M}_i, &T) \; \log_2 \bigg[\f{P(\mathcal{M}_i, T)}{P(\mathcal{M}_i, \t{RSS})} \bigg], \nonumber
\end{flalign}
shown in fig.~\rref{figure_2}{e}, where $D_\t{KL} \l[ P(\mathcal{M},T) \; || \; P(\mathcal{M},\t{RSS}) \r] \neq 0$ indicates how much more information the probability distribution $P(\mathcal{M},T)$ for the thermally equilibrated system contains in comparison to the random system $P(\mathcal{M},\t{RSS})$ due to the presence of SRO, i.e., eq.~\ref{eq:KL} quantifies the amount of SRO in the system. The close connection between information theory and thermodynamics\autocite{thermo_info,KL_correlation} makes it clear that a complete description of SRO also requires evaluating the spatial correlation\autocite{jim_null,jim_cem,correlated_disorder} among the motifs in addition to eq.~\ref{eq:KL}, which will be addressed later in this article.

Identifying the local chemical motif surrounding each atom also enables their association with atomic-scale physical properties. Consider, for example, that the chemical disorder of solid solutions distorts the perfect crystal lattice and leads to the creation of a heterogeneous landscape of local strain fields known as local lattice distortion. The association of chemical motifs with this characteristic property of solid solutions is shown in fig.~\rref{figure_2}{f}, where it can be seen that the local lattice distortion varies measurably with local chemical motifs. Notice how small differences in motifs can lead to substantial variations in physical properties: the two motifs connected in fig.~\rref{figure_2}{f} by the red arrow are related by the substitution of a single Cr atom by a Ni atom; this increases the associated local lattice distortion by 60\% and decreases the probability density from 40.7$\times$ more common than in a random solid solution to 3.5$\times$. This capability of capturing the subtle correlations between chemistry and physical properties is something that was lacking in the quantification of SRO, as demonstrated next.

\section{Warren-Cowley incompleteness}

The atom-centered characterization of chemical bond environments proposed in fig.~\rref{figure_2}{a} was motivated by the notoriously manybody nature of chemical bonds (i.e., not pairwise additive) and the electronic nearsightedness in condensed systems\autocite{nearsightedness}. Together, these two concepts suggest that a minimal representation requires $(n+1)$-body terms\autocite{N_body}, where $n$ is the central atom coordination number, such that the contribution of each nearest neighbor is accounted for on an equal footing. Here, this requirement is achieved in practice by including the connectivity among nearest neighbors in the graph representation of local chemical motifs (fig.~\rref{figure_2}{b}) and the message-passing algorithm of e3nn, which informs each graph node about the bonding topology of its neighbors. Representations based on \rev{first nearest-neighbor }WC parameters (eq.~\ref{eq:WC}) do not meet this requirement, consequently missing on important associations between local chemistry and physical properties \rev{due to the coupling between the WC description of neighboring atoms, which is properly included for atoms within the motif in our motif-based description}.

\begin{figure}[!bt]
  \centering
  \includegraphics{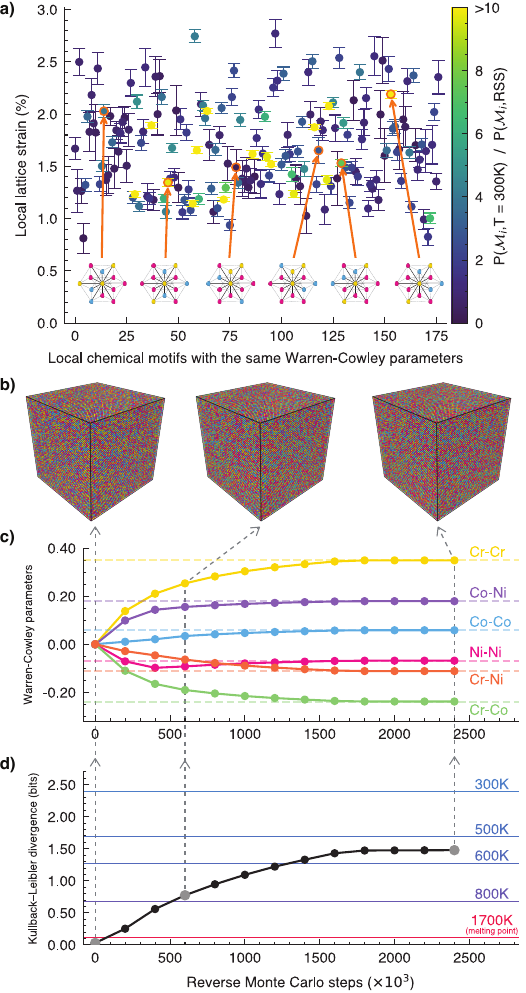}
  \caption{\textbf{Incompleteness of the WC representation.} \textbf{a)} Local lattice distortion and probability density at 300\,K for 182 local chemical motifs that are indistinguishable in WC-like representations. \textbf{b)} Illustration of the SRO evolution with a popular simulation approach that creates alloys with targeted WC parameters --- namely reverse Monte Carlo. \textbf{c)} Convergence of WC parameters towards the targeted thermal equilibrium values at 300\,K --- indicated by horizontal dashed lines. \textbf{d)} Horizontal lines are the equilibrium KL parameters (eq.~\ref{eq:KL} and fig.~\rref{figure_2}{e}) at the temperatures indicated on the right. The amount of SRO created is only 62\% of the SRO of a thermally equilibrated alloy because WC-like representations are not capable of capturing the entire complexity of local chemical motifs.}
  \label{figure_3}
\end{figure}

Consider for example the 182 unique motifs in which a Cr atom is surrounded by exactly seven Ni atoms, two Cr atoms, and three Co atoms. Within a \rev{first nearest-neighbor }WC-type of representation \rev{for the central atom,} all 182 motifs are considered equivalent as they contribute identically to eq.~\ref{eq:WC}. Yet, as shown in fig.~\rref{figure_3}{a}, these motifs display atomic-scale physical properties that vary substantially. For example, their probability density with respect to a random solid solution varies by two orders of magnitude while their local lattice distortion ranges from the 39th to the 97th percentile.

Another demonstration of the incompleteness of \rev{first nearest-neighbor }WC-like representations is shown in fig.~\rref{figure_3}{c}, where a reverse Monte Carlo method\autocite{RMC} is employed to generate alloys with targeted \rev{nearest-neighbor }WC parameters equivalent to those of a solid solution in thermal equilibrium at 300\,K. This is a popular approach employed in the investigation of SRO effects on microstructural evolution (e.g., refs.~\cite{OTIS} and \cite{flynn_PNAS}). However, fig.~\rref{figure_3}{d} shows that this approach does not create the correct amount of SRO: the final state has $D_\t{KL} = 1.48\;\t{bits}$, which corresponds to the SRO intensity at $550\t{K}$ (see fig.~\rref{figure_2}{e}) and is only 62\% of the actual $D_\t{KL} = 2.40\;\t{bits}$ for an alloy in thermal equilibrium at 300\,K. This discrepancy is due to the inability of \rev{first nearest-neighbor }WC-like representations to capture the entire complexity of local chemical motifs, which might lead to erroneous predictions of mechanical properties\autocite{curtin_mech} and thermodynamic stability\autocite{flynn_PNAS} (further illustrated in Supplementary Section 5).

\section{Short-range order length scale}
Quantification of the spatial organization of local chemical motifs is performed by evaluating the spatial correlation function $C_i(r,T)$ between a local chemical motif $\mathcal{M}_i$ and the motifs at a distance $r$ from $\mathcal{M}_i$. This correlation function is rigorously defined (see eq.~\ref{eq:C} and ref.~\cite{correlation}) with the assistance of a graph dissimilarity metric between $\mathcal{M}_i$ and other motifs $\mathcal{M}_j$: $d_{ij} = \lVert \v{d}_i -\v{d}_j \rVert_2$, which employs the embeddings $\v{d}_i$ obtained from e3nn. As shown in fig.~\rref{figure_2}{c} the space of dissimilarities among all motifs is rich in physical information on how the embeddings $\v{d}_i$ encode local chemical motifs (see also Supplementary Section 6 for more illustrations of the dissimilarity space).
\begin{figure}[!bt]
  \centering
  \includegraphics{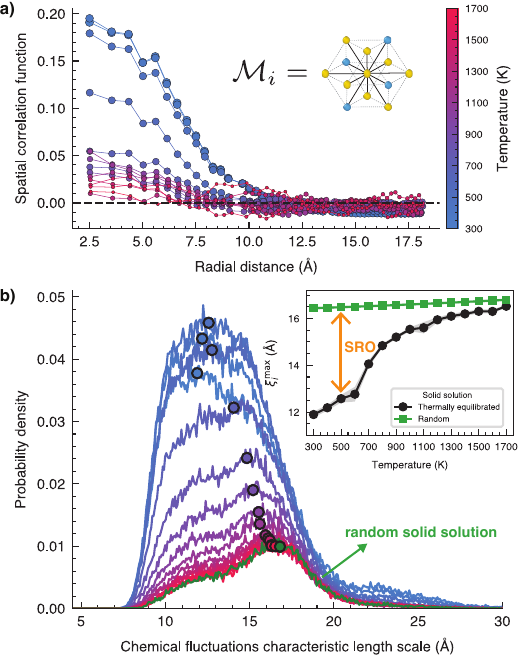}
  \caption{\textbf{Characteristic length-scale of chemical fluctuations}. \textbf{a)} Spatial correlation function of one of the two local chemical motifs $\mathcal{M}_i$ of a L1$_2$ ordered structure. \textbf{b)} Probability density of chemical fluctuations characteristic length scales $\xi_i$. Inset shows the temperature dependence of the maximum of the density distribution.}
  \label{figure_4}
\end{figure}

Figure \rref{figure_4}{a} shows the correlation function for one of the two motifs in a L1$_2$ ordered structure\autocite{mark_mrs,L12,L12_2,L12_3}. Other representative correlation functions are shown in Supplementary Section 7. One would prefer to have such information collected into a single metric, namely the \textit{characteristic length scale} ($\xi_i$) of chemical fluctuations. This can be realized by evaluating the radial distance at which motifs become uncorrelated, i.e., the shortest distance $r = r_i^*$ at which $C_i(r,T) = 0$ for $r \ge r_i^*$; the characteristic length scale is defined as twice that distance ($\xi_i = 2r_i^*$). The density distribution of $\xi_i$ (fig.~\rref{figure_4}{b}) reveals a complex dependence of the spatial extension of chemical fluctuations with temperature: the primary effect of increasing temperature is to decrease the total fraction of atoms displaying any short-range ordering. The significant presence of chemical fluctuations with magnitude as large as $25\,\Ang$ reinforces the importance of employing simulations at the appropriate length scale. This result is supported by the size-convergence analysis of WC parameters in Supplementary Section 1, which only converges for systems with dimensions larger than $\approx 25\,\Ang$.

The probability density of $\xi_i$ (fig.~\rref{figure_4}{b}) is central in quantifying the effect of SRO on materials properties. For example, it is a key parameter in the solute strengthening of high-entropy alloys\autocite{curtin_mech}. Figure \rref{figure_4}{b} can also be measured experimentally; yet, current efforts have been limited to capture $\xi_i$ for motifs of only a few ordered structures (such as L1$_1$\autocite{andy,ma_NiCoCr,ma} and L1$_2$\autocite{APT_CoCrNi,L12}). While electron diffraction measurements have being strongly contested\autocite{tem_challenge,resistivity_easo}, direct imaging\autocite{APT_CoCrNi} estimates $\xi_i \approx 20 \, \Ang$ for L1$_2$ motifs, which is in excellent agreement with our calculations (see Supplementary Section 8).

Finally, the distinction between SRO and random chemical fluctuations is made apparent in fig.~\rref{figure_4}{b} and its inset. Chemical fluctuations due to SRO become much more numerous than random fluctuations as temperature decreases, at the same time as their characteristic length scale decreases significantly below those of random fluctuations. Notice how this temperature trend is the opposite of what is observed during precipitation (including Guinier–Preston zones), where the length scale of the precipitating phase increases as temperature is decreased. The different temperature trend is due to the distinct nature of these two physical phenomena: precipitation requires the increase in number of a few motifs associated with a single ordered structure (e.g., two motifs for L1$_2$), meanwhile SRO involves many motifs (fig.~\rref{figure_2}{d}) with increasing population at the expense of less favorable motifs, turning the overall probability density $P(\mathcal{M},T)$ more sparse with decreasing temperature (fig.~\rref{figure_2}{e}), which leads to smaller length scales.

In summary, here we have presented a multiscale predictive approach to quantify the SRO state of metallic alloys with atomic resolution. \rev{While well-established cluster-based approaches are routinely employed to evaluate phase diagrams\autocite{S_config}, our framework focuses instead on enabling the comparison of atomistic simulations to various modern attempts to experimentally quantify the SRO length scale.} \rev{The results presented here} make it clear that a complete description of SRO requires the following two components. First, there is the tendency of certain chemical motifs $\mathcal{M}_i$ to be more common than others, which is fully characterized by the probability distribution $P(\mathcal{M},T)$ and can be summarized in a single SRO intensity metric --- namely the Kullback-Liebler divergence with respect to a random alloy. The second component quantifies how the motifs are organized in space, which is fully characterized by correlation functions $C_i(r,T)$ and can be summarized in a single characteristic length scale metric $\l<\xi_i\r>$. Our findings advance the foundational understanding of solid-solution phases and help place current experimental efforts in characterizing SRO into a physically rigorous framework that is independent of features and limitations of particular experimental techniques.

\clearpage
\twocolumn[
  \begin{center}
	\large
	\textbf{Methods}
  \end{center}
]

\section{Machine learning interatomic potential}
The Moment Tensor Potential\autocite{MTP} was employed as the model for ML-IAPs. Identical training procedures and hyperparameters were employed for both ML-IAPs (i.e., with and without chemical sampling): $R_\t{cut} = 5\,\Ang$ (corresponding to a cutoff between the 3rd and 4th coordination shells), $\t{lev}_\t{max} = 20$ (corresponding to 651 independent model parameters), a Chebyshev radial basis with size of $8\,\Ang$, energy data weight of 1.0, force data weight of 0.01, and a maximum of 10\,000 iterations using the Broyden–Fletcher–Goldfarb–Shannon algorithm with an error tolerance of $10^{-5}$.

A complete description of all training and test sets is available in the Supplementary Section 3. Below we provide a short summary of these data sets.

The training set for the ML-IAP without chemical sampling was inspired by the popular approach introduced in refs.~\cite{SPO_MLIAP_1} and \cite{SPO_MLIAP_2} for NbMoTaW. This training set contained a total of 83\,970 atoms spread across various snapshots including: single-element ground states and distorted structures, surface slabs with different orientations, and ab initio molecular dynamics simulations at different temperatures (including above the melting point), random binary systems with various compositions, and Special Quasi-random Structures (SQS\autocite{SQS_1,SQS_2}) of ternary ab initio molecular dynamics simulations at different temperatures (including above the melting point).

The training set for the ML-IAP with chemical sampling contained a total of 108\,684 atoms spread across various snapshots. Half of these atoms are from snapshots in the crystalline phase with chemical ordering extracted from various points along DFT Monte Carlo simulations. Random thermal noise\autocite{dc3} of various magnitudes were added to these snapshots, which were also isotropically expanded to account for thermal expansion effects. The other half of the snapshots were extracted from a single ab initio molecular dynamics simulation at 1800\,K (i.e., right above the melting point $T_\t{m} = 1690\t{K}$\autocite{Tm} for CrCoNi). An ensemble of five independent ML-IAPs were produced with this data set in order to establish that model differences due to random initialization are smaller than our precision in measuring them\autocite{bagging,deep_learning_book,ensemble_shapeev}. The best performing model from the ensemble was chosen based on a model averaging approach\autocite{deep_learning_book}. 

The data in the test sets shown in fig.~\rref{figure_1}{b} were obtained from snapshots and thermodynamic conditions not employed in the training of either ML-IAP potentials. The ``Random solid solution'' test set included only random CrCoNi systems with various degrees of thermal noise and thermal expansion. The ``Thermally equilibrated solid solution'' test set included chemical ordering extracted along DFT Monte Carlo simulations at 750K and 1200K with thermal noise and thermal expansion. The ``Liquid'' snapshots were extracted from ab initio molecular dynamics simulations at 2684K.

\section{Density-functional theory}
All DFT calculations were performed using the Vienna ab initio simulation package (VASP\autocite{VASP_1,VASP_2,VASP_3,VASP_4,VASP_PAW}) version 6.2.1 with the Perdew–Burke–Ernzerhof\autocite{PBE} exchange-correlation functional (version \texttt{potpaw\_PBE.54}) and projector-augmented plane wave\autocite{PAW} potentials. Pseudo-potentials were employed such that the valence electrons of each element are $3\t{p}^6 3\t{d}^5 4\t{s}^1$ for Cr, $3\t{d}^8 4\t{s}^1$ for Co, and $3\t{d}^8 4\t{s}^2$ for Ni. All DFT calculations (including Monte Carlo and molecular dynamics) are spin-polarized (collinear) with initial magnetic-moment configurations of 0.6\,$\mu_\t{B}$ for Cr, 2.0\,$\mu_\t{B}$ for Co, and 1.0\,$\mu_\t{B}$ for Ni --- where $\mu_\t{B}$ is the Bohr magneton.

Convergence tests were performed to determine the following parameters: energy cutoff of 430\,eV, second-order Methfessel-Paxton smearing with a sigma value of 0.1\,eV, and K-point grid of $6 \times 6 \times 6$ for a face-centered cubic unit cell, which was scaled proportionally for larger symmetric systems. The DFT Monte Carlo and molecular dynamics simulations were carried out with a single $\Gamma$ K-point, but the snapshots employed in the training or test of ML-IAPs had their energies recomputed with a K-point grid of equivalent density to the symmetric systems in order to maintain the same precision. An energy threshold of $10^{-5}\,\t{eV}$ was employed in all self-consistent loops, and a force threshold of $0.02\,\t{eV}/\Ang$ was employed for all structural relaxations. Ab initio molecular dynamics simulations employed a Langevin thermostat with a friction coefficient $\gamma = 10\,\t{ps}^{-1}$.

All structural manipulations and analyses of DFT calculations were performed using Ovito\autocite{ovito}, Python Materials Genomics\autocite{pymatgen}, and Fireworks\autocite{fireworks}.

\section{Monte Carlo}
Monte Carlo simulations for thermally equilibrated solid solutions with respect to SRO were performed using the Metropolis-Hasting algorithm. All of the Monte Carlo simulations in fig.~\rref{figure_1}{a} were identical except for the energy calculation method (DFT, IAP, or ML-IAPs): five independent simulations at 500\,K using a $3\times 3\times 3$ supercell with 108 atoms were performed with 5000 atom-swap attempts carried out for each. The WC parameters in fig.~\rref{figure_1}{a} were computed using the last 100 snapshots from these simulations. Meanwhile, a total of 101 snapshots evenly distributed over the course of each DFT Monte Carlo simulation were extracted for training ML-IAPs with chemical sampling.

The results in figs.~\ref{figure_2}, \ref{figure_3}, and \ref{figure_4} were obtained from 216 independent and size-converged (see Supplementary Section 1) Monte Carlo simulations with 4000 atoms, where 30 atom-swap attempts were performed per atom (60\,000 total attempts). These simulations were repeated from room temperature (300\,K) to the melting point (1700\,K) in intervals of 100\,K. Notice that while the probability density $P(\mathcal{M}, T)$ for finite temperatures $T$ is estimated from the Monte Carlo simulations, the probability $P(\mathcal{M}, \t{RSS})$ for a random solid solution can be evaluated exactly by creating a data set of all possible $3^{13} = 1\,594\,323$ possible chemical motifs.

The final Monte Carlo configurations were structurally relaxed (i.e., energy minimized) with fixed cell size and employed in the calculation of the local lattice distortion for each atom $n$, defined as
\begin{equation}
  \label{eq:lattice_distortion}
  \delta_n(T) = \frac{\l\lVert \v{r}_n^{\t{f}} - \v{r}_n^{i} \r\rVert_2}{a_\t{NN}(T)} ,
\end{equation}  
where $\v{r}_n^\t{f}$ is the final position after the structural relaxation, $\v{r}_n^\t{i}$ is the initial position in the ideal face-centered cubic structure (accounting for thermal expansion), $\lVert \ldots \rVert_2$ denotes the $L^2$ norm, and $a_\t{NN}(T)$ is the nearest-neighbor distance at temperature $T$ (the temperature dependence comes from the thermal expansion of the lattice).

\section{$\mathbb{E}$(3)-invariant embedding}
Local chemical motif $\mathcal{M}_i$ (fig.~\rref{figure_2}{a}) was extracted from Monte Carlo snapshots and transformed into an independent graph $g(\mathcal{M}_i)$ representation (fig.~\rref{figure_2}{b}) where each node represents an atom. The graph nodes have the one-hot encoded atomic type as an attribute, while the graph edges store the distance vector between the nodes connected by them. \rev{Invariance of the graph embedding with respect to local lattice distortions is accomplished by mapping atoms back to their ideal lattice positions before the construction of the graphs}. This graph takes the form of a cuboctahedron for the face-centered cubic lattice, with 8 triangular faces, 6 square faces, and 12 vertices, representing the 12 first-nearest neighbors surrounding the central atom. This adds up to 36 identical edges, with 12 of them connecting the central atom to its first neighbors, and the other 24 connecting first neighbors to each other (as shown in fig.~\rref{figure_2}{b}).

Each graph $g(\mathcal{M}_i)$ is passed through an $\mathbb{E}$(3)-equivariant graph neural network using the e3nn package\autocite{e3nn_2} for implementing E(3) neural networks. The neural network architecture\autocite{e3nn_3} consists of two \rev{$\mathbb{E}(3)$}-equivariant convolutions \rev{composed of $\mathcal{O}(3)$-equivariant filters} with spherical harmonics $Y_\ell^m(r)$ up to degree $\ell_\t{max}=1$ added as edge attributes, ten cosine radial basis functions evenly spread out over the range from zero to $2.25\times$ the nearest neighbor distance, and an output length of 100 scalars collected into a vector $\v{z}_i\in\mathbb{R}^{100}$. The data in $\mathbb{E}$(3) neural networks implemented with e3nn are irreducible representations typed by their angular frequency $\ell$ and parity (even or odd), for example a vector transforms as the irreducible representation with angular frequency $\ell=1$ and odd parity is denoted in e3nn as `$1o$'. For the hidden-layers representation of the convolution (in the e3nn syntax) we features that transform as the irreducible representations $50\times1o+50\times1e+50\times0o+50\times0e$. The neural network is randomly initialized in order to maximize the influence of each weight --- an approach that is supported by the use of such networks as symmetry compilers capable of representing geometrical features\autocite{e3nn_4}.

\section{Reverse Monte Carlo}
The reverse Monte Carlo\autocite{RMC} simulations of figs.~\rref{figure_3}{b}, \rref{figure_3}{c}, and \rref{figure_3}{d} employed the Metropolis-Hasting algorithm applied to reverse-engineer atomic configurations that obey the observed WC parameters $\alpha^\t{target}_\t{AB}$ at 300\,K. The probability of acceptance of each atom-swap move is $\min \l[1, \exp(-\Delta \chi^2/ \sigma^2) \r]$, where $\sigma^2 = 10^{-9}$ and $\Delta \chi^2 = \chi_\t{f}^2 - \chi_\t{i}^2$ with 
\[
  \chi_\t{i}^2 = \sum\limits_{AB} (\alpha^\t{target}_\t{AB} - \alpha^\t{i}_\t{AB})^2
\]
quantifying how far a configuration $i$ is from the targeted WC parameters. $\chi_\t{f}$ is the same quantity after an attempted atom swap move. In Supplementary Section 5 we demonstrate similar results at various other temperatures, including employing an alternative strategy\autocite{OTIS} for reverse-engineering WC parameters.

\section{Dissimilarity metric}

The dissimilarity metric between two local chemical motifs $\mathcal{M}_i$ and $\mathcal{M}_j$ is defined as $d_{ij} = \lVert \v{d}_i - \v{d}_j \rVert_2$, where $\lVert \ldots \rVert_2$ denotes the $L^2$ norm. The vector $\v{d}_i \in \mathbb{R}^5$ \rev{was designed to respect geometric symmetries and} can be decomposed into three orthogonal vectors (see Supplementary Section 6 for more details) with simple physical and geometrical interpretations: $\v{d}_i = \f{1}{5} \v{c}_i + \f{2}{5} \v{k}_i + \f{2}{5} \v{s}_i$ such that
\begin{equation}
  \label{eq:d}
  d_{ij} = \f{1}{5} \, \lVert \v{c}_i - \v{c}_j \rVert_2 + \f{2}{5} \, \lVert \v{k}_i - \v{k}_j \rVert_2 + \f{2}{5} \, \lVert \v{s}_i - \v{s}_j \rVert_2.
\end{equation}
The central-atom vector ($\v{c}_i$) is such that $\lVert \v{c}_i - \v{c}_j \rVert_2 = 1$ if the central atom of both motifs have the same chemical type, and zero otherwise. The chemical composition vector ($\v{k}_i$) of the central atom 12 nearest neighbors is such that $\big\lVert \v{k}_i - \v{k}_j \big\rVert_2 = 1$ for motifs at different vertices of the ternary concentration triangle, and $\v{s}_i$ is the projection of the output vector of e3nn $\v{z}_i \in \mathbb{R}^{100}$ along the direction of largest variance among all motifs with the same concentration $\v{k}_i$ as $\mathcal{M}_i$ normalized such that $0 \leq \big\lVert \v{s}_i - \v{s}_j \big\rVert_2 \leq 1$. Equation \ref{eq:d} results in the dissimilarity space illustrated in fig.~\rref{figure_2}{c}.

The terms in eq.~\ref{eq:d} have a simple interpretation: $\lVert \v{c}_i - \v{c}_j \rVert_2$ encodes the dissimilarity between motifs with different central atoms, $\lVert \v{k}_i - \v{k}_j \rVert_2$ encodes different chemical concentration of the central atom nearest neighbors (this term can be loosely connected to WC parameters), and $\lVert \v{s}_i - \v{s}_j \rVert_2$ encodes the dissimilarity between motifs that contribute equally to WC parameters but have different local chemical bond environment. Each term in $\v{d}_i$ (and eq.~\ref{eq:d}) is weighted proportionally to the number of bonds (i.e., edges) affected by its corresponding structure in fig.~\rref{figure_2}{b}: 12 for the central atom and 24 for the motif overall composition, and 24 for the motif structural configuration. The overall normalization is such that $0\leq d_{ij} \leq 1$, where $d_{ij} = 0$ if and only if $\mathcal{M}_i = \mathcal{M}_j$.

Illustrations of the geometrical meaning of $\v{c}_i$, $\v{k}_i$, $\v{s}_i$, and the dissimilarity space are available in the Supplementary Section 6 and Supplementary Video 1.

\section{Correlation function}

With the dissimilarity metric of eq.~\ref{eq:d} one can rigorously define (see eq.~9 in ref.~\cite{correlation}) the correlation function $C_i(r,T)$ of a motif $\mathcal{M}_i$ with other motifs located at a radial distance $r$ from $\mathcal{M}_i$:
\begin{equation}
  \label{eq:C}
  C_i(r,T) = \phi_i(r,T)  - \phi^0_i(r,T)
\end{equation}
where
\begin{equation}
  \label{eq:phi_i}
  \phi_i(r,T) = 1 - 2 \, \big< d_{ij} \big>_{|\v{r}_i-\v{r}_j| = r} 
\end{equation}
with $\l< \ldots \r>_{|\v{r}_i-\v{r}_j| = r}$ indicating an average including only motifs $\mathcal{M}_j$ (located at $\v{r}_j$) at a distance $r$ from $\mathcal{M}_i$ (located at $\v{r}_i$), and
\[
  \phi^0_i(r,T) = 1 - 2 \, \big< d_{ij} \big>_{P(\mathcal{M},T)}
\]
where $\l< \ldots \r>_{P(\mathcal{M},T)}$ indicates an average evaluated with motifs $\mathcal{M}_j$ randomly sampled from the thermally equilibrated distribution $P(\mathcal{M},T)$. The two terms in eq.~\ref{eq:C} are evaluated as follows. First, $\phi_i(r,T)$ is computed using the 216 Monte Carlo simulations with 4000 atoms at each temperature $T$. Meanwhile, $\phi_i^0(r,T)$ is computed as the weighted average of the dissimilarities of $\mathcal{M}_i$ with all other motifs, with the weights being assigned based on $P(\mathcal{M},T)$. The probability $P(\mathcal{M},T)$ itself is estimated from the Monte Carlo simulations. Notice the existence of geometrically-necessary correlations between $\mathcal{M}_i$ and the motifs of the neighbors of the central atom in $\mathcal{M}_i$ up to the fourth coordination shell, which all share atoms with $\mathcal{M}_i$. For these four coordination shells $\phi_i^0(r,T)$ is evaluated by first building the probability distribution $P_{i,n}(\mathcal{M},T)$ of all geometrically-compatible motifs of $\mathcal{M}_i$ in the $n$th coordination shell observed in the Monte Carlo simulations and then computing the dissimilarity only between $\mathcal{M}_i$ and its geometrically compatible motifs. $\phi_i^0(r, T)$ is then determined as a weighted average of the dissimilarities, with the weights being assigned based on $P_{i,n}(\mathcal{M},T)$.

The correlation function in eq.~\ref{eq:C} is such that $C_i(r\rightarrow \infty,T) = 0$ because $\phi^0_i(r,T)$ is the correlation function between $\mathcal{M}_i$ and an uncorrelated distribution of motifs drawn from the same thermally equilibrated distribution used to evaluate $\phi_i(r,T)$. The radial distance at which motif $\mathcal{M}_i$ becomes uncorrelated is the shortest distance $r_i^*$ such that $C_i(r,T) = 0$ for $r \ge r_i^*$. The characteristic length scale (used to construct the histogram in fig.~\rref{figure_4}{b}) is then defined as twice that distance $\xi_i = 2r_i^*$. See Supplementary Section 7 for a complete description of the statistical analysis of $C_i(r,T)$ and also other representative examples such as fig.~\rref{figure_4}{a}.

\section{Data and code availability}

\rev{The software for chemical motif identification and SRO quantification can be found in our \texttt{ChemicalMotifIdentifier} Python package\autocite{cmi_github}. The potential can be found in our \texttt{MachineLearningPotential} GitHub repository\autocite{potentialgithub}. Our figure style is implemented in \texttt{LovelyPlots}\autocite{lovelyplots} under the \texttt{paper} style. For convenience, we have compiled the list of all Python packages developed in this work in a GitHub repository list\autocite{repo_list}. 
Any custom code or data that is not currently available in these repositories can be subsequently added upon reasonable request to the corresponding author.}

\section{Author contributions}

K.S., Y.C., T.S., and R.F. conceived the project.
Y.C. performed all DFT and Monte Carlo simulations, and the ML-IAP training and validation.
K.S. performed all calculations to characterize and quantify SRO through local chemical motifs, including the reverse Monte Carlo simulations. 
All authors contributed to the interpretation of the results. 
K.S., Y.C., and R.F. prepared the manuscript, which was reviewed and edited by all authors.
Project administration, supervision, and funding acquisition was performed by R.F.

\section{Acknowledgments}

This work was supported by the MathWorks Ignition Fund, MathWorks Engineering Fellowship Fund, and the Portuguese Foundation for International Cooperation in Science,
Technology and Higher Education in the MIT--Portugal Program. We were also supported by the Research Support Committee Funds from the School of Engineering at the Massachusetts Institute of Technology. This work used the Expanse supercomputer at the San Diego Supercomputer Center through allocation MAT210005 from the Advanced Cyberinfrastructure Coordination Ecosystem: Services \& Support (ACCESS) program, which is supported by National Science Foundation grants \#2138259, \#2138286, \#2138307, \#2137603, and \#2138296, and the Extreme Science and Engineering Discovery Environment (XSEDE), which was supported by National Science Foundation grant number \#1548562.

\section{Competing interests}

The authors declare no competing interests.

\clearpage
\printbibliography[heading=bibnumbered,title={References}]

@article{curtin_mech,
  title={Theory of strengthening in fcc high entropy alloys},
  author={Varvenne, C{\'e}line and Luque, Aitor and Curtin, William A.},
  journal={Acta Materialia},
  volume={118},
  pages={164--176},
  year={2016},
  publisher={Elsevier},
  doi={10.1016/j.actamat.2016.07.040}
}

@article{dc3,
  title={Data-centric framework for crystal structure identification in atomistic simulations using machine learning},
  author={Chung, Heejung W. and Freitas, Rodrigo and Cheon, Gowoon and Reed, Evan J.},
  journal={Physical Review Materials},
  volume={6},
  number={4},
  pages={043801},
  year={2022},
  publisher={APS},
  doi={10.1103/PhysRevMaterials.6.043801}
}

@article{exafs,
  title={Local structure and short-range order in a NiCoCr solid solution alloy},
  author={Zhang, F. X. and Zhao, Shijun and Jin, Ke and Xue, H. and Velisa, G. and Bei, H. and Huang, R. and Ko, J. Y. P. and Pagan, D. C. and Neuefeind, J. C. and Weber,  W. J. and Zhang, Y.},
  journal={Physical Review Letters},
  volume={118},
  number={20},
  pages={205501},
  year={2017},
  publisher={APS},
  doi={10.1103/PhysRevLett.118.205501}
}

@article{finnis,
  title={Magnetism and thermodynamics of defect-free Fe-Cr alloys},
  author={Klaver, T. P. C. and Drautz, R. and Finnis, M. W.},
  journal={Physical Review B},
  volume={74},
  number={9},
  pages={094435},
  year={2006},
  publisher={APS},
  doi={10.1103/PhysRevB.74.094435}
}

@article{doug_jim,
  title={Spin-driven ordering of Cr in the equiatomic high entropy alloy NiFeCrCo},
  author={Niu, C. and Zaddach, A. J. and Oni, A. A. and Sang, X. and Hurt, J. W. and LeBeau, J. M. and Koch, C. C. and Irving, D. L.},
  journal={Applied Physics Letters},
  volume={106},
  number={16},
  year={2015},
  publisher={AIP Publishing},
  doi={10.1063/1.4918996}
}

@article{jim_cem,
  title={Determination of short-range order in TiVNbHf (Al)},
  author={Xu, Michael and Wei, Shaolou and Tasan, C. Cem and LeBeau, James M.},
  journal={Applied Physics Letters},
  volume={122},
  number={18},
  year={2023},
  publisher={AIP Publishing},
  doi={10.1063/5.0145289}
}

@article{penghui,
  title={Maximum strength and dislocation patterning in multi--principal element alloys},
  author={Cao, Penghui},
  journal={Science Advances},
  volume={8},
  number={45},
  pages={eabq7433},
  year={2022},
  publisher={American Association for the Advancement of Science},
  doi={10.1126/sciadv.abq7433}
}

@article{ma_NiCoCr,
  title={Atomic-scale evidence of chemical short-range order in CrCoNi medium-entropy alloy},
  author={Zhou, Lingling and Wang, Qi and Wang, Jing and Chen, Xuefei and Jiang, Ping and Zhou, Hao and Yuan, Fuping and Wu, Xiaolei and Cheng, Zhiying and Ma, En},
  journal={Acta Materialia},
  volume={224},
  pages={117490},
  year={2022},
  publisher={Elsevier},
  doi={10.1016/j.actamat.2021.117490}
}

@article{tamm,
  title={Atomic-scale properties of Ni-based FCC ternary, and quaternary alloys},
  author={Tamm, Artur and Aabloo, Alvo and Klintenberg, Mattias and Stocks, Malcolm and Caro, Alfredo},
  journal={Acta Materialia},
  volume={99},
  pages={307--312},
  year={2015},
  publisher={Elsevier},
  doi={10.1016/j.actamat.2015.08.015}
}

@article{flynn_PNAS,
  title={Magnetically driven short-range order can explain anomalous measurements in CrCoNi},
  author={Walsh, Flynn and Asta, Mark and Ritchie, Robert O.},
  journal={Proceedings of the National Academy of Sciences},
  volume={118},
  number={13},
  pages={e2020540118},
  year={2021},
  publisher={National Academy Sciences},
  doi={10.1073/pnas.2020540118}
}

@article{andy,
  title={Short-range order and its impact on the CrCoNi medium-entropy alloy},
  author={Zhang, Ruopeng and Zhao, Shiteng and Ding, Jun and Chong, Yan and Jia, Tao and Ophus, Colin and Asta, Mark and Ritchie, Robert O. and Minor, Andrew M.},
  journal={Nature},
  volume={581},
  number={7808},
  pages={283--287},
  year={2020},
  publisher={Nature Publishing Group UK London},
  doi={10.1038/s41586-020-2275-z}
}

@article{ma,
  title={Direct observation of chemical short-range order in a medium-entropy alloy},
  author={Chen, Xuefei and Wang, Qi and Cheng, Zhiying and Zhu, Mingliu and Zhou, Hao and Jiang, Ping and Zhou, Lingling and Xue, Qiqi and Yuan, Fuping and Zhu, Jing and  Wu, Xiaolei and Ma, En},
  journal={Nature},
  volume={592},
  number={7856},
  pages={712--716},
  year={2021},
  publisher={Nature Publishing Group UK London},
  doi={10.1038/s41586-021-03428-z}
}

@article{SRO_radiation,
  title={Effect of local chemical order on the irradiation-induced defect evolution in CrCoNi medium-entropy alloy},
  author={Zhang, Zhen and Su, Zhengxiong and Zhang, Bozhao and Yu, Qin and Ding, Jun and Shi, Tan and Lu, Chenyang and Ritchie, Robert O. and Ma, Evan},
  journal={Proceedings of the National Academy of Sciences},
  volume={120},
  number={15},
  pages={e2218673120},
  year={2023},
  publisher={National Academy of Sciences},
  doi={10.1073/pnas.2218673120}
}

@article{SRO_indirect,
  title={Dependence of electrical resistivity on the degree of short range order in a nickel--copper alloy},
  author={Wagner, W. and Poerschke, R. and Wollenberger, H.},
  journal={Philosophical Magazine B},
  volume={43},
  number={2},
  pages={345--355},
  year={1981},
  publisher={Taylor \& Francis},
  doi={10.1080/13642818108221904}
}

@article{WC_complete,
  title={Atomic short-range order and incipient long-range order in high-entropy alloys},
  author={Singh, Prashant and Smirnov, Andrei V. and Johnson, Duane D.},
  journal={Physical Review B},
  volume={91},
  number={22},
  pages={224204},
  year={2015},
  publisher={APS},
  doi={10.1103/PhysRevB.91.224204}
}

@article{diffuse_origin,
  title={On the origin of diffuse intensities in fcc electron diffraction patterns},
  author={Coury, Francisco Gil and Miller, Cody and Field, Robert and Kaufman, Michael},
  journal={Nature},
  volume={622},
  number={7984},
  pages={742--747},
  year={2023},
  publisher={Nature Publishing Group UK London},
  doi={10.1038/s41586-023-06530-6}
}

@article{APT_CoCrNi,
  title={Direct observation of local chemical ordering in a few nanometer range in CoCrNi medium-entropy alloy by atom probe tomography and its impact on mechanical properties},
  author={Inoue, Koji and Yoshida, Shuhei and Tsuji, Nobuhiro},
  journal={Physical Review Materials},
  volume={5},
  number={8},
  pages={085007},
  year={2021},
  publisher={APS},
  doi={10.1103/PhysRevMaterials.5.085007}
}

@article{SRO_SFE,
  title={Tunable stacking fault energies by tailoring local chemical order in CrCoNi medium-entropy alloys},
  author={Ding, Jun and Yu, Qin and Asta, Mark and Ritchie, Robert O.},
  journal={Proceedings of the National Academy of Sciences},
  volume={115},
  number={36},
  pages={8919--8924},
  year={2018},
  publisher={National Academy of Sciences},
  doi={10.1073/pnas.1808660115}
}

@article{fontaine,
  title={The number of independent pair-correlation functions in multicomponent systems},
  author={de Fontaine, Didier},
  journal={Journal of Applied Crystallography},
  volume={4},
  number={1},
  pages={15--19},
  year={1971},
  publisher={International Union of Crystallography},
  doi={10.1107/S0021889871006174}
}

@book{khachaturyan,
  title={Theory of Structural Transformations in Solids},
  author={Khachaturyan, Armen G.},
  year={2008},
  publisher={Dover}
}

@article{SRO_dislocations,
  title={Strengthening in multi-principal element alloys with local-chemical-order roughened dislocation pathways},
  author={Li, Qing-Jie and Sheng, Howard and Ma, Evan},
  journal={Nature Communications},
  volume={10},
  number={1},
  pages={3563},
  year={2019},
  publisher={Nature Publishing Group UK London},
  doi={10.1038/s41467-019-11464-7}
}

@article{WC_1,
  title={An approximate theory of order in alloys},
  author={Cowley, John M.},
  journal={Physical Review},
  volume={77},
  number={5},
  pages={669},
  year={1950},
  publisher={APS},
  doi={10.1103/PhysRev.77.669}
}

@book{WC_2,
  title={X-ray Diffraction},
  author={Warren, Bertram Eugene},
  year={1990},
  publisher={Courier Corporation}
}

@article{SPO_MLIAP_1,
  title={Complex strengthening mechanisms in the NbMoTaW multi-principal element alloy},
  author={Li, Xiang-Guo and Chen, Chi and Zheng, Hui and Zuo, Yunxing and Ong, Shyue Ping},
  journal={npj Computational Materials},
  volume={6},
  number={1},
  pages={70},
  year={2020},
  publisher={Nature Publishing Group UK London},
  doi={10.1038/s41524-020-0339-0}
}

@article{SPO_MLIAP_2,
  title={Multi-scale investigation of short-range order and dislocation glide in   MoNbTi and TaNbTi multi-principal element alloys},
  author={Zheng, Hui and Fey, Lauren TW and Li, Xiang-Guo and Hu, Yong-Jie and Qi, Liang and Chen, Chi and Xu, Shuozhi and Beyerlein, Irene J. and Ong, Shyue Ping},
  journal={npj Computational Materials},
  volume={9},
  number={1},
  pages={89},
  year={2023},
  publisher={Nature Publishing Group UK London},
  doi={10.1038/s41524-023-01046-z}
}

@article{SPO_MLIAP_performance,
  title={Performance and cost assessment of machine learning interatomic potentials},
  author={Zuo, Yunxing and Chen, Chi and Li, Xiangguo and Deng, Zhi and Chen, Yiming and Behler, Jörg and Csányi, Gábor and Shapeev, Alexander V. and Thompson, Aidan P. and Wood, Mitchell A. and Ong, Shyue Ping},
  journal={The Journal of Physical Chemistry A},
  volume={124},
  number={4},
  pages={731--745},
  year={2020},
  publisher={ACS Publications},
  doi={10.1021/acs.jpca.9b08723}
}

@article{polya,
  title={{Kombinatorische Anzahlbestimmungen für Gruppen, Graphen und chemische Verbindungen}},
  author={George P{\'o}lya},
  journal={Acta Mathematica},
  volume={68},
  number={none},
  pages={145 -- 254},
  year={1937},
  publisher={Institut Mittag-Leffler},
  doi={10.1007/BF02546665},
}

@article{e3nn_1,
  title={Euclidean symmetry and equivariance in machine learning},
  author={Smidt, Tess E.},
  journal={Trends in Chemistry},
  volume={3},
  number={2},
  pages={82--85},
  year={2021},
  publisher={Elsevier},
  doi={10.1016/j.trechm.2020.10.006}
}

@misc{e3nn_2,
  title={{e3nn: Euclidean Neural Networks}}, 
  author={Geiger, Mario and Smidt, Tess},
  year={2022},
  archivePrefix={arXiv},
  publisher={arXiv},
  doi={10.48550/arXiv.2207.09453},
}

@article{tfn,
  title={Tensor field networks: Rotation-and translation-equivariant neural networks for 3d point clouds},
  author={Thomas, Nathaniel and Smidt, Tess and Kearnes, Steven and Yang, Lusann and Li, Li and Kohlhoff, Kai and Riley, Patrick},
  journal={arXiv},
  year={2018},
  doi={https://doi.org/10.48550/arXiv.1802.08219}
}

@article{cgn,
  title={Clebsch--Gordan nets: a fully fourier space spherical convolutional neural network},
  author={Kondor, Risi and Lin, Zhen and Trivedi, Shubhendu},
  journal={Advances in Neural Information Processing Systems},
  volume={31},
  year={2018},
  doi={https://proceedings.neurips.cc/paper/2018/hash/a3fc981af450752046be179185ebc8b5-Abstract.html}
}

@article{cnn,
  title={3d steerable cnns: Learning rotationally equivariant features in volumetric data},
  author={Weiler, Maurice and Geiger, Mario and Welling, Max and Boomsma, Wouter and Cohen, Taco S},
  journal={Advances in Neural Information Processing Systems},
  volume={31},
  year={2018},
  doi={https://proceedings.neurips.cc/paper_files/paper/2018/hash/488e4104520c6aab692863cc1dba45af-Abstract.html}
}

@article{L12_2,
  title={Chemical domain structure and its formation kinetics in CrCoNi medium-entropy alloy},
  author={Du, Jun-Ping and Yu, Peijun and Shinzato, Shuhei and Meng, Fan-Shun and Sato, Yuji and Li, Yangen and Fan, Yiwen and Ogata, Shigenobu},
  journal={Acta Materialia},
  volume={240},
  pages={118314},
  year={2022},
  publisher={Elsevier},
  doi={10.1016/j.actamat.2022.118314}
}

@article{L12_3,
  title={Short-range order and phase stability of CrCoNi explored with machine learning potentials},
  author={Ghosh, Sheuly and Sotskov, Vadim and Shapeev, Alexander V. and Neugebauer, J{\"o}rg and K{\"o}rmann, Fritz},
  journal={Physical Review Materials},
  volume={6},
  number={11},
  pages={113804},
  year={2022},
  publisher={APS},
  doi={10.1103/PhysRevMaterials.6.113804}
}

@article{L12,
  title={Data-driven electron-diffraction approach reveals local short-range ordering in CrCoNi with ordering effects},
  author={Hsiao, Haw-Wen and Feng, Rui and Ni, Haoyang and An, Ke and Poplawsky, Jonathan D. and Liaw, Peter K. and Zuo, Jian-Min},
  journal={Nature Communications},
  volume={13},
  number={1},
  pages={6651},
  year={2022},
  publisher={Nature Publishing Group UK London},
  doi={10.1038/s41467-022-34335-0}
}

@article{N_body,
  title={Incompleteness of atomic structure representations},
  author={Pozdnyakov, Sergey N. and Willatt, Michael J. and Bart{\'o}k, Albert P. and Ortner, Christoph and Cs{\'a}nyi, G{\'a}bor and Ceriotti, Michele},
  journal={Physical Review Letters},
  volume={125},
  number={16},
  pages={166001},
  year={2020},
  publisher={APS},
  doi={10.1103/PhysRevLett.125.166001}
}

@article{WL_1,
  title={The reduction of a graph to canonical form and the algebra which appears therein},
  author={Weisfeiler, Boris and Lehman, Andrei A.},
  journal={Nauchno-Technicheskaya Informatsia},
  volume={2},
  number={9},
  pages={12--16},
  year={1968}
}

@article{MB_WC,
  title={Quantitative description of atomic architecture in solid solutions: a generalized theory for multicomponent short-range order},
  author={Ceguerra, Anna V. and Powles, Rebecca C. and Moody, Michael P. and Ringer, Simon P.},
  journal={Physical Review B},
  volume={82},
  number={13},
  pages={132201},
  year={2010},
  publisher={APS},
  doi={10.1103/PhysRevB.82.132201}
}

@article{S_config,
  title={Cluster expansion of alloy theory: a review of historical development and modern innovations},
  author={Kadkhodaei, Sara and Mu{\~n}oz, Jorge A.},
  journal={JOM},
  volume={73},
  number={11},
  pages={3326--3346},
  year={2021},
  publisher={Springer},
  doi={10.1007/s11837-021-04840-6}
}

@inproceedings{WL_2,
  title={How Powerful are Graph Neural Networks?},
  author={Keyulu Xu and Weihua Hu and Jure Leskovec and Stefanie Jegelka},
  booktitle={International Conference on Learning Representations},
  year={2019},
  doi={10.48550/arXiv.1810.00826}
}

@article{nearsightedness,
  title={Nearsightedness of electronic matter},
  author={Prodan, Emil and Kohn, Walter},
  journal={Proceedings of the National Academy of Sciences},
  volume={102},
  number={33},
  pages={11635--11638},
  year={2005},
  publisher={National Academy of Sciences},
  doi={10.1073/pnas.0505436102}
}

@article{correlation,
  title={Constructing correlation coefficients from similarity and dissimilarity functions},
  author={Batyrshin, Ildar Z.},
  journal={Acta Polytechnica Hungarica},
  volume={16},
  number={10},
  pages={191--204},
  year={2019},
  publisher={Budapest Tech Polytechnic Institution},
  doi={10.12700/APH.16.10.2019.10.12}
}

@article{tem_challenge,
  title={Extra electron reflections in concentrated alloys do not necessitate short-range order},
  author={Walsh, Flynn and Zhang, Mingwei and Ritchie, Robert O. and Minor, Andrew M. and Asta, Mark},
  journal={Nature Materials},
  volume={22},
  number={8},
  pages={926--929},
  year={2023},
  publisher={Nature Publishing Group UK London},
  doi={10.1038/s41563-023-01570-9}
}

@article{jim_null,
  title={Correlating local chemical and structural order using Geographic Information Systems-based spatial statistics},
  author={Xu, Michael and Kumar, Abinash and LeBeau, James M.},
  journal={Ultramicroscopy},
  volume={243},
  pages={113642},
  year={2023},
  publisher={Elsevier},
  doi={10.1016/j.ultramic.2022.113642}
}

@article{OTIS,
  title={Random generation of lattice structures with short-range order},
  author={Fey, Lauren T. W. and Beyerlein, Irene J.},
  journal={Integrating Materials and Manufacturing Innovation},
  volume={11},
  number={3},
  pages={382--390},
  year={2022},
  publisher={Springer},
  doi={10.1007/s40192-022-00269-0}
}

@book{KL,
  title={Elements of Information Theory},
  author={Cover, Thomas M.},
  year={1999},
  publisher={John Wiley \& Sons}
}

@article{thermo_info,
  title={Thermodynamics of information},
  author={Parrondo, Juan M. R. and Horowitz, Jordan M. and Sagawa, Takahiro},
  journal={Nature Physics},
  volume={11},
  number={2},
  pages={131--139},
  year={2015},
  publisher={Nature Publishing Group UK London},
  doi={10.1038/nphys3230}
}

@article{KL_correlation,
  title={Network information and connected correlations},
  author={Schneidman, Elad and Still, Susanne and Berry, Michael J. and Bialek, William},
  journal={Physical Review Letters},
  volume={91},
  number={23},
  pages={238701},
  year={2003},
  publisher={APS},
  doi={10.1103/PhysRevLett.91.238701}
}

@article{RMC,
  title={Reverse Monte Carlo modelling},
  author={McGreevy, Robert L.},
  journal={Journal of Physics: Condensed Matter},
  volume={13},
  number={46},
  pages={R877},
  year={2001},
  publisher={IOP Publishing},
  doi={10.1088/0953-8984/13/46/201}
}

@article{EXAFS_SRO,
  title={Why is EXAFS for complex concentrated alloys so hard? Challenges and opportunities for measuring ordering with X-ray absorption spectroscopy},
  author={Joress, Howie and Ravel, Bruce and Anber, Elaf and Hollenbach, Jonathan and Sur, Debashish and Hattrick-Simpers, Jason and Taheri, Mitra L and DeCost, Brian},
  journal={Matter},
  year={2023},
  publisher={Elsevier},
  doi={10.1016/j.matt.2023.09.010}
}

@article{mark_mrs,
  title={Reconsidering short-range order in complex concentrated alloys},
  author={Walsh, Flynn and Abu-Odeh, Anas and Asta, Mark},
  journal={MRS Bulletin},
  volume={48},
  number={7},
  pages={753--761},
  year={2023},
  publisher={Springer},
  doi={10.1557/s43577-023-00555-y}
}

@article{resistivity_easo,
  title={Evolution of short-range order and its effects on the plastic deformation behavior of single crystals of the equiatomic Cr-Co-Ni medium-entropy alloy},
  author={Li, Le and Chen, Zhenghao and Kuroiwa, Shogo and Ito, Mitsuhiro and Yuge, Koretaka and Kishida, Kyosuke and Tanimoto, Hisanori and Yu, Yue and Inui, Haruyuki and George, Easo P.},
  journal={Acta Materialia},
  volume={243},
  pages={118537},
  year={2023},
  publisher={Elsevier},
  doi={10.1016/j.actamat.2022.118537}
}

@article{HEC_2,
  title={Cation-disordered rocksalt-type high-entropy cathodes for Li-ion batteries},
  author={Lun, Zhengyan and Ouyang, Bin and Kwon, Deok-Hwang and Ha, Yang and Foley, Emily E and Huang, Tzu-Yang and Cai, Zijian and Kim, Hyunchul and Balasubramanian, Mahalingam and Sun, Yingzhi and Huang, J. and Tian, Y. and Kim, H. and McCloskey, B. D. and Yang, W. and Clément, R. J. and Ji, H. and Ceder, G.},
  journal={Nature Materials},
  volume={20},
  number={2},
  pages={214--221},
  year={2021},
  publisher={Nature Publishing Group UK London},
  doi={10.1038/s41563-020-00816-0}
}

@article{HEC_3,
  title={Probing the local site disorder and distortion in pyrochlore high-entropy oxides},
  author={Jiang, Bo and Bridges, Craig A. and Unocic, Raymond R. and Pitike, Krishna Chaitanya and Cooper, Valentino R. and Zhang, Yuanpeng and Lin, De-Ye and Page, Katharine},
  journal={Journal of the American Chemical Society},
  volume={143},
  number={11},
  pages={4193--4204},
  year={2020},
  publisher={ACS Publications},
  doi={10.1021/jacs.0c10739}
}

@article{SRO_simulation,
  title={Simulating short-range order in compositionally complex materials},
  author={Ferrari, Alberto and K{\"o}rmann, Fritz and Asta, Mark and Neugebauer, J{\"o}rg},
  journal={Nature Computational Science},
  volume={3},
  number={3},
  pages={221--229},
  year={2023},
  publisher={Nature Publishing Group US New York},
  doi={10.1038/s43588-023-00407-4}
}

@article{HEC_1,
  title={Atomic-resolution electron microscopy of nanoscale local structure in lead-based relaxor ferroelectrics},
  author={Kumar, Abinash and Baker, Jonathon N. and Bowes, Preston C. and Cabral, Matthew J. and Zhang, Shujun and Dickey, Elizabeth C. and Irving, Douglas L. and LeBeau, James M.},
  journal={Nature Materials},
  volume={20},
  number={1},
  pages={62--67},
  year={2021},
  publisher={Nature Publishing Group UK London},
  doi={10.1038/s41563-020-0794-5}
}

@article{correlated_disorder,
  title={The crystallography of correlated disorder},
  author={Keen, David A. and Goodwin, Andrew L.},
  journal={Nature},
  volume={521},
  number={7552},
  pages={303--309},
  year={2015},
  publisher={Nature Publishing Group UK London},
  doi={10.1038/nature14453}
}

@article{MTP,
  title={Moment tensor potentials: A class of systematically improvable interatomic potentials},
  author={Shapeev, Alexander V.},
  journal={Multiscale Modeling \& Simulation},
  volume={14},
  number={3},
  pages={1153--1173},
  year={2016},
  publisher={SIAM},
  doi={10.1137/15M1054183}
}

@article{Tm,
  title={Temperature dependence of the mechanical properties of equiatomic solid solution alloys with face-centered cubic crystal structures},
  author={Wu, Zhenggang and Bei, Hongbin and Pharr, George M. and George, Easo P.},
  journal={Acta Materialia},
  volume={81},
  pages={428--441},
  year={2014},
  publisher={Elsevier},
  doi={10.1016/j.actamat.2014.08.02}
}

@article{VASP_1,
  title={Ab initio molecular dynamics for liquid metals},
  author={Kresse, Georg and Hafner, J{\"u}rgen},
  journal={Physical Review B},
  volume={47},
  number={1},
  pages={558},
  year={1993},
  publisher={APS},
  doi={10.1103/PhysRevB.47.558}
}

@article{VASP_2,
  title={Efficient iterative schemes for ab initio total-energy calculations using a plane-wave basis set},
  author={Kresse, Georg and Furthm{\"u}ller, J{\"u}rgen},
  journal={Physical Review B},
  volume={54},
  number={16},
  pages={11169},
  year={1996},
  publisher={APS},
  doi={10.1103/PhysRevB.54.11169}
}

@article{VASP_3,
  title={Efficiency of ab-initio total energy calculations for metals and semiconductors using a plane-wave basis set},
  author={Kresse, Georg and Furthm{\"u}ller, J{\"u}rgen},
  journal={Computational Materials Science},
  volume={6},
  number={1},
  pages={15--50},
  year={1996},
  publisher={Elsevier},
  doi={10.1016/0927-0256(96)00008-0}
}

@article{VASP_4,
  title={Ab initio molecular-dynamics simulation of the liquid-metal-amorphous-semiconductor transition in germanium},
  author={Kresse, Georg and Hafner, J{\"u}rgen},
  journal={Physical Review B},
  volume={49},
  number={20},
  pages={14251},
  year={1994},
  publisher={APS},
  doi={10.1103/PhysRevB.49.14251}
}

@article{VASP_PAW,
  title={From ultrasoft pseudopotentials to the projector augmented-wave method},
  author={Kresse, Georg and Joubert, Daniel},
  journal={Physical Review B},
  volume={59},
  number={3},
  pages={1758},
  year={1999},
  publisher={APS},
  doi={10.1103/PhysRevB.59.1758}
}

@article{PBE,
  title={Generalized gradient approximation made simple},
  author={Perdew, John P. and Burke, Kieron and Ernzerhof, Matthias},
  journal={Physical Review Letters},
  volume={77},
  number={18},
  pages={3865},
  year={1996},
  publisher={APS},
  doi={10.1103/PhysRevLett.77.3865}
}

@article{PAW,
  title={Projector augmented-wave method},
  author={Bl{\"o}chl, Peter E.},
  journal={Physical Review B},
  volume={50},
  number={24},
  pages={17953},
  year={1994},
  publisher={APS},
  doi={10.1103/PhysRevB.50.17953}
}

@article{ovito,
  title={{Visualization and analysis of atomistic simulation data with OVITO-the Open Visualization Tool}},
  author={Stukowski, Alexander},
  journal={Modelling and Simulation in Materials Science and Engineering},
  volume={18},
  number={{1}},
  year={2010},
  doi={10.1088/0965-0393/18/1/015012}
}

@article{pymatgen,
  title={{Python Materials Genomics (pymatgen): A robust, open-source python library for materials analysis}},
  author={Ong, Shyue Ping and Richards, William Davidson and Jain, Anubhav and Hautier, Geoffroy and Kocher, Michael and Cholia, Shreyas and Gunter, Dan and Chevrier, Vincent L. and Persson, Kristin A. and Ceder, Gerbrand},
  journal={Computational Materials Science},
  volume={68},
  pages={314--319},
  year={2013},
  publisher={Elsevier},
  doi={10.1016/j.commatsci.2012.10.028}
}

@article{fireworks,
  title={{FireWorks}: a dynamic workflow system designed for high-throughput applications},
  author={Jain, Anubhav and Ong, Shyue Ping and Chen, Wei and Medasani, Bharat and Qu, Xiaohui and Kocher, Michael and Brafman, Miriam and Petretto, Guido and Rignanese, Gian-Marco and Hautier, Geoffroy and Gunter, Daniel and Persson, Kristin A.},
  journal={Concurrency and Computation: Practice and Experience},
  volume={27},
  number={17},
  pages={5037--5059},
  year={2015},
  publisher={Wiley Online Library},
  doi={10.1002/cpe.3505}
}

@article{e3nn_3,
  title={A recipe for cracking the quantum scaling limit with machine learned electron densities},
  author={Rackers, Joshua A. and Tecot, Lucas and Geiger, Mario and Smidt, Tess E},
  journal={Machine Learning: Science and Technology},
  volume={4},
  number={1},
  pages={015027},
  year={2023},
  publisher={IOP Publishing},
  doi={10.1088/2632-2153/acb314}
}

@article{e3nn_4,
  title={Finding symmetry breaking order parameters with {E}uclidean neural networks},
  author={Smidt, Tess E. and Geiger, Mario and Miller, Benjamin Kurt},
  journal={Physical Review Research},
  volume={3},
  number={1},
  pages={L012002},
  year={2021},
  publisher={APS},
  doi={10.1103/PhysRevResearch.3.L012002}
}

@article{SQS_1,
  title={Special quasirandom structures},
  author={Zunger, Alex and Wei, S. H. and Ferreira, L. G. and Bernard, James E.},
  journal={Physical Review Letters},
  volume={65},
  number={3},
  pages={353},
  year={1990},
  publisher={APS},
  doi={10.1103/PhysRevLett.65.353}
}

@article{HEA,
  title={High-entropy alloys},
  author={George, Easo P. and Raabe, Dierk and Ritchie, Robert O.},
  journal={Nature Reviews Materials},
  volume={4},
  number={8},
  pages={515--534},
  year={2019},
  publisher={Nature Publishing Group UK London},
  doi={10.1038/s41578-019-0121-4}
}

@article{SQS_2,
  title={{Efficient stochastic generation of Special Quasirandom Structures}},
  author={A. van de Walle and P. Tiwary and M. M. de Jong and D. L. Olmsted and M. D. Asta and A. Dick and D. Shin and Y. Wang and L.-Q. Chen and Z.-K. Liu},
  journal={Calphad},
  volume={42},
  pages={13--18},
  year={2013},
  doi={10.1016/j.calphad.2013.06.006}
}

@article{ensemble_shapeev,
  title={Impact of lattice relaxations on phase transitions in a high-entropy alloy studied by machine-learning potentials},
  author={Kostiuchenko, Tatiana and K{\"o}rmann, Fritz and Neugebauer, J{\"o}rg and Shapeev, Alexander},
  journal={npj Computational Materials},
  volume={5},
  number={1},
  pages={55},
  year={2019},
  publisher={Nature Publishing Group UK London},
  doi={10.1038/s41524-019-0195-y}
}

@article{bagging,
  title={Bagging predictors},
  author={Breiman, Leo},
  journal={Machine learning},
  volume={24},
  pages={123--140},
  year={1996},
  publisher={Springer}
}

@book{deep_learning_book,
  title={Deep Learning},
  author={Ian Goodfellow and Yoshua Bengio and Aaron Courville},
  publisher={MIT Press},
  note={\url{http://www.deeplearningbook.org}},
  year={2016}
}

@misc{xu2019powerful,
      title={How Powerful are Graph Neural Networks?}, 
      author={Keyulu Xu and Weihua Hu and Jure Leskovec and Stefanie Jegelka},
      year={2019},
      eprint={1810.00826},
      archivePrefix={arXiv},
      primaryClass={cs.LG}
}

@software{figures,
  url={https://github.com/killiansheriff/figures_quantifying_sro_in_metallic_alloys},
  author={Sheriff, Killian and Cao, Yifan and Freitas, Rodrigo},
}

@software{lovelyplots,
  author={Sheriff, Killian},
  title={LovelyPlots: A collection of matplotlib stylesheets for scientific figures},
  month=aug,
  year=2023,
  publisher={Zenodo},
  doi={10.5281/zenodo.6903936},
  url={https://github.com/killiansheriff/LovelyPlots}
}

@software{repo_list,
  url={https://github.com/stars/killiansheriff/lists/quantifying-sro-in-alloys},
  author={Sheriff, Killian and Cao, Yifan and Freitas, Rodrigo},
}

@software{cmi_github,
	author       = {Sheriff, Killian and Cao, Yifan and Freitas, Rodrigo},
	url          = {https://github.com/killiansheriff/ChemicalMotifIdentifier}
}

@software{potentialgithub,
  url={https://github.com/yifan-henry-cao/MachineLearningPotential},
  author={Cao, Yifan and Sheriff, Killian and Freitas, Rodrigo},
}

\end{document}